# Microservice-based edge platform for AI services

P. Lalanda, G. Vega and D. Morand
LIG Laboratory,
Université Grenoble Alpes
France

*Abstract*—Pervasive computing promotes the integration of smart electronic devices in our living and working spaces to provide advanced services. Recently, two major evolutions are changing the way pervasive applications are developed. The first deals with moving computation and storage to the edge. The second is the massive use of machine learning techniques to build these applications. However, architectural principles and integrated frameworks are still missing today to successfully and repetitively support application developers in the creation of edge-level AI applications. In this paper, we present a novel architecture and platform allowing the development of such applications in smart spaces.

## I. INTRODUCTION

Pervasive computing promotes the integration of smart electronic devices in our living and working spaces to provide a wide variety of services. This concept is now implemented in many fields and a number of applications have been developed and deployed [2]. A recent trend is the use of more decentralized architectures where computation is done near the data sources. This approach, referred to as edge computing [9], allows for better resource management and security. Another major evolution is the use of Machine-Learning (ML) techniques. The goal of an ML system is to train an algorithm to automatically make decisions by identifying patterns that may be hidden within massive data sets whose exact nature is unknown and cannot be programmed explicitly.

The use of ML techniques opens the way for more advanced applications on the edge. But the development of these new applications raises formidable challenges that are not addressed by traditional software engineering techniques. ML-based applications are not developed in the same way as more traditional applications. They are built, tested, installed, configured, run, monitored and updated differently. Because of that, and despite impressive performances, they are actually very difficult to deploy in the real world.

The purpose of this paper is to define the life cycle of the new ML-based components, show what are the implied new requirements, and present a new pervasive platform, based on microservices, adapted to the execution and administration of ML-based services on the edge. This platform is an evolution of the iCasa platform and is validated on a real industrial use case. The paper is organized as follows. First, some background and clarification on ML-based pervasive applications is provided. Then our use case is introduced. Section IV is about our proposal and section V details the implementation of that solution. Finally, the article concludes with a discussion of the results obtained and a projection on future work.

## II. MACHINE LEARNING ON THE EDGE

Two main workflows are set up to build ML models. The first workflow, called model development, is primarily performed by data scientists. This workflow assumes that a business problem has been properly identified along with sources of historical data. During the first steps, raw historical data is analyzed, cleaned and transformed into appropriate numeric representations called features. Feature engineering is a complex task which purpose is to find out the most relevant data representations given the available data, the task at hand and the targeted model. Features are then used to train a model with a carefully chosen machine learning algorithm. Once a model has been developed, it is deployed on a target execution machine. This is the essence of the second workflow that is often implemented by software engineers. Its purpose is first to collect appropriate data and make predictions. As the name suggests, predictions are only ... predictions. They are founded on data which has not been seen during the training phase and can therefore be incorrect. Finally, collected data might be set aside in order to be used for further training.

The prediction model is a major artifact of a software systems based on learning techniques. This model has its own life cycle, as illustrated by figure 2. The initial development uses historical data; its relevance depends on the quality of the available data and adequacy of the selected learning algorithm. In the case of pervasive applications, this data must come from the field, or from very high quality simulations, to be usable. This is an iterative process where the steps of data collection, feature selection and model training are repeated until a satisfactory result is obtained.

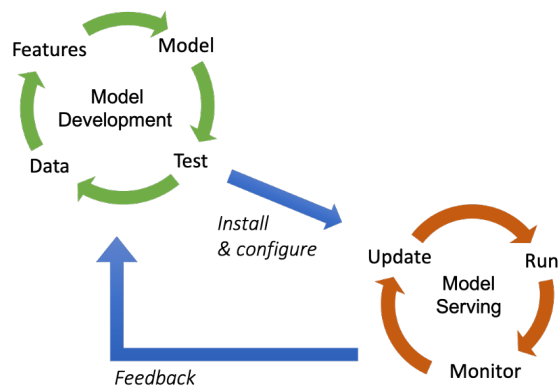

Fig. 1: Life-cycle of ML-based components

Model deployment activities look like traditional deployment tasks but are nevertheless significantly different. First of all, the model is transferred to the edge machine and installed like any other software artifact. It is also configured to fit properly into its execution environment. Configuration can be complex and actually depends on lot on the services provided by the host middleware. In some cases, the model has to be further trained with local data in order to adjust to the specificity of the execution environment. Then appropriate data has to be collected in order to run the model, which can thus provide the expected predictions. The model is constantly monitored by system administrators, through appropriate tools. Finally, an update step completes the loop. The purpose of this step is either to update the model directly on site or to send data back to the data scientist teams.

A model is updated when a new one has been devised by data scientists, which happens fairly often. This can be done to enhance its performance, to improve its compactness, or even to extend its functional scope. Sometimes, this also comes with a change in the input data, which may require updates on data gathering activities. Finally, let us note that the models are extremely sensitive to data evolution, even very slight one. The smallest modification can invalidate a prediction model. Therefore, they must be continuously adapted to the execution environment and the corresponding data to stay relevant.

We believe that these particularities require the development of a new generation of pervasive platforms with advanced features, better able to handle the massive data requirements and the particular life cycle of prediction models. Precisely, we believe that such platform should meet the following requirements:

- Deployment support. In our target architecture, models are transferred to the edge from cloud servers. Communication and deployment facilities are needed to support such transfer activities in secure and efficient ways.
- Data collection. Collecting data in pervasive environment requires to deal with heterogeneous protocols but also to make choices on the networks to use, on the data to collect, how to collect them in order to properly feed the machine learning based services.
- Model execution support. There are today a number of machine learning frameworks, generally based on different languages. A pervasive platform should be able to host models built in different languages in a transparent way.
- Model retraining support. As explained before, the notions of retraining and continual learning are of major importance in Industry 4.0. It is important that applications run on a pervasive platform can continue their training in order to improve themselves and get personalized.
- Model monitoring support. It is of major importance to monitor the performance of ML-based service, which may degrade because of a changing environment or inappropriate data. Mechanisms are then needed to allow the detection of such deviations.

III. USE CASE

HVAC systems are the subject of important research activities in Industry 4.0, essentially because of their cost in terms of infrastructure support, asset management and energy. In that regard, chillers are the most energy consuming components of large size buildings. The amount of electricity consumed by a chiller is actually not only determined by the total cooling load but also by its energy-efficiency (which also differs depending on their cooling regime). Intuitively, if this efficiency is low, then more electricity is consumed to support a required cooling demand. It is regularly necessary to determine the best way to activate chillers depending on their efficiency.

Precisely, operating the most efficient combination of chillers in a building in real-time in order to meet time-varying cooling demands is called chiller sequencing. For example, sequencing a building with two chillers [0.5, 0.7] implies that chiller 1 and chiller 2 are operating at 50 percent and 70 percent of their maximum rated capacity, respectively. The sequencing problem is to allocate the cooling load at any given time to the chillers in the most energy efficient manner so that the overall cooling demand of the building is satisfied while at the same time the electricity consumed by the chillers is kept at a minimum [7]. The efficacy of chiller sequencing control relies heavily on the run-time performance profile of the chillers, namely the COP (Coefficient of performance) under different cooling load regimes. COP is a measure of the energy efficiency of a chiller and captures the cooling power that it can output for a certain input power consumption. The cooling demand changes over time, so chiller sequencing must be performed repeatedly in order to continuously meet the varying cooling demand. To ensure cooling performance, chiller sequencing needs time for feedback control until the system regains stability when switching from one sequence to another. The chiller sequencing for each period must be completed before the start of the next sequencing period.

A usual solution is to use manufacturer values for the COP. These values correspond to chillers performance profile when they leave the factory. Although, it corresponds to the current state-of-the-art, it is limited in the sense that, rapidly, default COP values do not accurately reflect reality. COP are indeed very sensitive to chillers aging and to the running conditions. Recent work proposed to compute individualized COP for chillers by applying machine learning techniques using historical chiller data [11] [10]. Precisely, a private cloud is established to store the historical data from a BMS (Building Management System) and to train a prediction model. When a cooling demand arrives, the cloud can perform chiller sequencing assisted by a model-based COP prediction. This approach has demonstrated its interest. The performance of the solution has been evaluated by applying it to BMS data, spanning 4 years, obtained from multiple chillers across three large commercial buildings in Hong Kong. It has been showed that the proposed solution can save over 30 percent of HVAC electricity consumption.

As illustrated by figure 3, we have designed an architecture where computing and storage functions are distributed more effectively between cloud and fog. Simply put, COP models for chillers are trained in a cloud infrastructure and sent to a fog infrastructure where they can be run and retrained with data collected on the plant floor. Updated models, using new data or new ML techniques, are regularly computed at the cloud level and sent to the fog where they have to be integrated. In such architecture, pervasive services (in blue) do not replace existing control systems (in red), often implemented with programmable logic controllers. The goal is rather to provide complementary services, based on secondary sensing, with relaxed demands regarding real-time requirements.

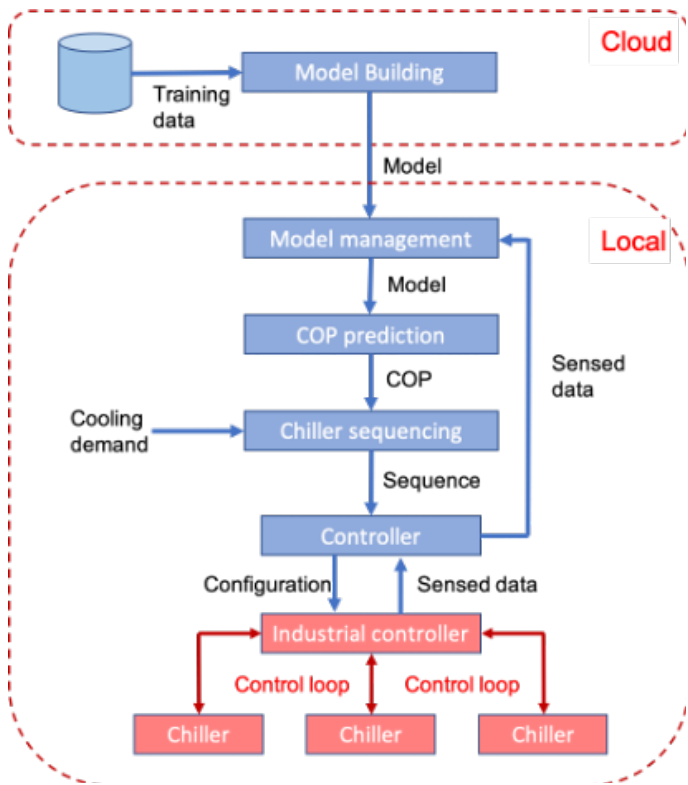

Fig. 2: Life-cycle of ML-based components

This architecture has been successfully implemented on top of the iCasa pervasive platform [6]. Nevertheless, this work confirmed a set of problems (as noted above):

- It is not able to deal with massive flows of data. Current platforms, like iCasa, build a context that stores a limited amount of data and are not able to deal with all the data needed for the learning process.
- The integration of Python-based model in other programming languages, like Java in our case, is difficult. This does not allow for regular and automated updates. The integration requires specific code that is difficult, if not impossible, to generate.
- It does not allow the easy integration of the asynchronous delivery of different teams

## IV. Our approach

We have defined a pervasive platform based on the microservice paradigm and designed for the edge. It allows the development and execution of context-aware applications that can be autonomically adapted at runtime [5]. Microservices allow the definition of component-based system where components are implemented as loosely coupled units (aptly called microservices) that can be independently deployed, managed, and updated. To achieve this, each microservice is packaged as a self-contained deployment and execution unit. The architecture is presented by Figure 4. It is made of four major microservices implemented with the docker technology:

- A *Device access manager* and *Context* module collect data from devices or services and present it in an appropriate form to the applications run on the gateway. The context also reifies devices so that they can be modified by the applications.
- A *Time series database* allows longer term storage of collected data with a time stamp. This database is suitable for temporal queries.
- A *Machine Learning manager* deals with prediction models, including their call, update and needed data.
- *Applications* are developed and run on top of these modules that must be customized to specific environments.

The device access manager [1] allows to access data provided by pervasive devices. It supports an open set of protocols, including Zwave, Zigbee, X10, UPnP, DPWS and Bluetooth. The principle of this manager is to generate proxies allowing to interact in a transparent way with the devices. and generates proxy services: when a device or a remote service disappears, the manager detects the departure and removes the proxy. The set of managed protocols can be extended at runtime by adding a protocol manager, using generic facilities. The device access manager is also in charge of monitoring concurrent invocations. Most of the protocols do not support concurrent accesses. To enforce integrity, read and write accesses are made sequential. Usage quotas and fair-use are also automatically enforced.

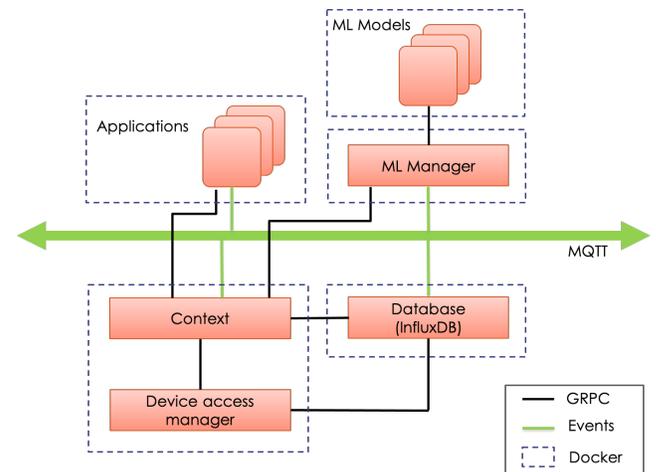

Fig. 3: Microservice-based gateway

The device access manager is implemented with iPOJO, a service-oriented development model [3] ], which in is based on the OSGi framework (see www.osgi.org). The iPOJO development model allows the definition of software components defined by their name, the specifications of provided and required services, their properties and, of course, an implementation class (Java). Required services are supplemented with cardinalities. Also, they can be optional or mandatory, substitutable at runtime or not, and constrained by predicates (first order logic applied to component properties). Bindings are done and updated dynamically at runtime by the execution framework.

The platform also includes a context module using the device access manager to get data from devices or to trigger an action on those devices. Its goal is to dynamically present contextual information captured in the environment. Such information is presented through gRPC APIs (Google Remote Procedure Call) or through events sent to the interested applications. Context is dynamic in order to reflect the changing nature of the execution environment but also to deal with applications evolving needs. In fact, the context receives the data needs of the applications and must provide them in the best possible way, according to the devices and therefore the available data. For this, the device access manager may have to use remote platforms [8]. The context is also implemented with ipojo. This is why context and device access manager are kept in the same microservice. These two components are integrated seamlessly and provide a number of built-in functions that ease the programmer work.

A time-series database containing time-stamped data has also been integrated. Data is generally simple and unstructured. Specifically, our platform integrates Influx DB which turned out to be very efficient to store unstructured IoT-based data, indexed by time. The database is used to store all the data collected from devices. It is heavily used by the context and order to get time-stamped information but also to store structured information that it builds. With this architectural structure, the context only presents information required by the running application. When the needed information evolves, the context can be refreshed with historical data kept in the database. The database also publishes data to the ML framework. This data corresponds to the features needed by the ML models and can also change over time.

We have defined a specific module to manage the different machine-learning models hosted by the platform. This module provides gRPC interfaces to call a model and also to get its characteristics. The module is also able to dynamically subscribe to data needed by the models and provided by the Influx Database. Data thus received are used to make predictions but also for the model updates, which may be done locally or on the cloud. According to our architecture, models are executed as specific micro services. They are deployed dynamically and characterized by a set of metadata including the needed input data. It is also to be noted that several models can be used for a single prediction (redundancy pattern) and it is then up the ML manager to decide on the prediction to use (with a vote mechanism for instance).

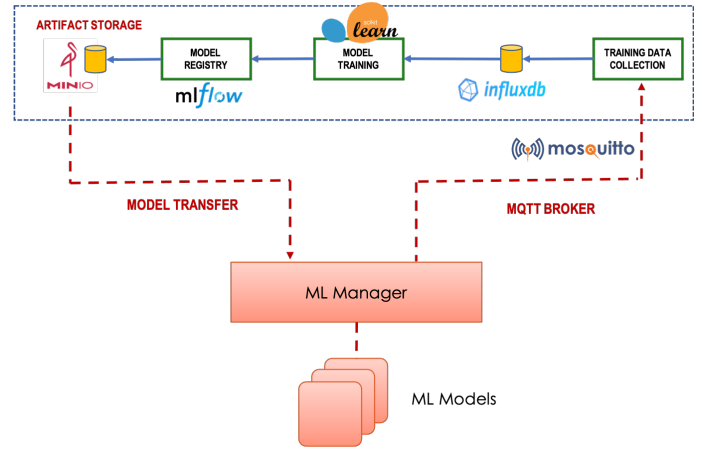

Fig. 4: Cloud architecture

In the cloud, the learning workflow is implemented and supported by a number of tools (see figure 5). First of all, data collected in the plant floor are retrieved with Mosquitto MQTT. This is a lightweight middleware suitable for use on all devices from low power single board computers to full servers and then ideal for pervasive settings. Collected data is then stored in the Influx Database, just like at the gateway level. Then the Scikit Learn framework is used to train the models with the newly received data (added to historical ones). We integrated an open source platform to store the different model versions. MLflow Registry offers a centralized model store to manage the computed models (and above all its different versions related to the used dataset used for training). It includes model lineage (which MLflow experiment and run produced the model)and model versioning capabilities.

## V. APPLICATION IMPLEMENTATION

### A. Initial model

The initial training of the COP prediction model is performed in the cloud. As stated before, this initial training is based on the successful work presented in section 4. It includes the global approach and the different algorithms selected for this study, as explained in []. The total data collected from the BMS is more than 1 TB. A private cloud has been configured in order to process the data for all the experiments, with 16 cores of 2.6GHz CPU and a total memory of 64GB. Models have been trained with three-year data and their accuracy (F1-score) has been tested with one-year data, which is a common setting in time-series data mining and multi-task learning.

Technically speaking, we used the scikit-learn framework and several algorithms for comparison purposes, including linear regression, Support Vector Regression (SVR), and AdaBoost. It clearly appeared that the ensemble approach like AdaBoost (1) can better capture the non-linearity than linear regression, and (2) are less likely to become over-fitted than Support Vector Regression on large datasets, due to the model combination nature of AdaBoost.

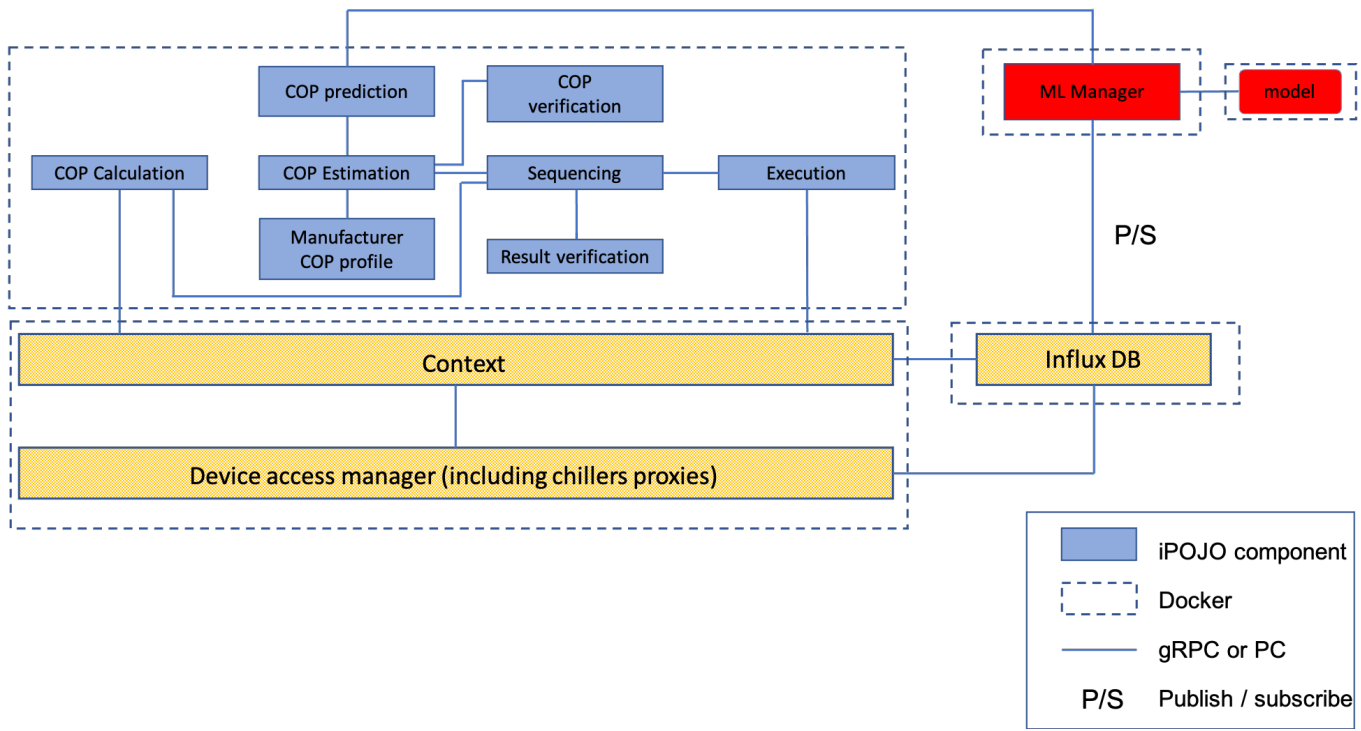

Fig. 5: Application implementation

## B. Application architecture

The application core is implemented in iPOJO and is then based on the service-oriented paradigm. The software architecture is in figure 4 and is made of the following components:

- The *Sequencing* component is the main software component. It orchestrates the whole application. It receives a cooling demand and decides on the right configuration to be applied to each chiller and the sequence of activation of those chillers.
- The *COP Estimation* component is the first to be called by the *Sequencing* component. It provides the current COP. To do so, it can use either a prediction provided by the *COP Predictor* component or a more conservative value corresponding to the *Manufacturer COP profile* component. In our system, this latter solution is only used when the prediction is rejected by the *Data verification* component.
- The *Manufacturer COP profile* component provides an extrapolation of the COP based on the chiller's manufacturer model profile. The provided value is in general not very accurate but safe since it remains within an acceptable range of values. As explained before, this is the approach used in most systems.
- The *COP prediction* component provides an estimation of the COP based on machine-learning techniques. To do so, it calls the *ML manager*, which in turn calls the model deployed from the cloud, based on AdaBoost. The model uses data provided by the Influx database including the season, the age of the chillers, the external temperature, the weather characterization, the chiller brand and model, the chiller power (kw), the water mass flow, the temperature difference between the in and out water, the last cooling demand, etc.
- The *COP verification* component checks that the COP estimation is in an acceptable range of values and coherent with the previous values. The purpose of this component is to eliminate grossly false values. It does not have the ability to detect small errors.
- The *Result verification* component checks that the sequencing is coherent with the previous sequence of activation. Just like the previous component, this component only allows to detect major deviations.
- The *COP computation* component calculates the COP with some delay. It does so indirectly through measures in the environment. This COP corresponds to a previous sequencing cycle and is calculated *a posteriori* when the system regains stability. It is used to label feature vectors previously collected and then allows future training.
- The *Execution* component interacts with the chillers through the context in order to implement the computed activations.

The global behavior of this architecture is the following. Data is collected every second and stored in the database. Important information (current and past) is presented as services by the context through appropriate APIs. When a cooling demand is set, a new COP is calculated using the machine-learning model and, then, chillers sequencing is computed. If the sequencing is rejected by the verification component, a more conservative COP based on a simple extrapolation is used.

Let us note that bindings can be dynamically changed at runtime: this allows updating many aspects of the application without stopping it. In particular, the verification components and the ML components can be changed anytime. Thanks to the microservice paradigm, components updates are managed independently. We are also working on having several ML models at the same time so that the best one can be selected according to the execution context.

Regarding performance, between 500 and 1000 measures can be collected every second. A similar number of items is also written in the database every second. This range is conservative: we checked that the architecture could support ten times more data. The amount of data sent back to the cloud is of course way smaller. The ML model is relatively small and is executed in real time. The complete cycle is executed in well under a second, which is well within the requirements (of the order of a few seconds).

*C. Discussion*

The platform presented in this paper has greatly facilitated the implementation of the chiller use case. Let us revisit some of the requirements that were highlighted in Section 3 and see how they are met.

**Deployment**. This is made easy by the docker technology that has been selected to implement the microservices. As explained, updates are done dynamically and the life cycle of each microservice is managed independently. This brings a lot of flexibility and also allows each of the actors of the project (software engineers, business analysts, data scientists) to develop in their own language and at their own pace.

**Data collection**. This aspect is taken care of by the context and the device access manager. These two frameworks are quite complete (for example, many protocols are already present) and easily configurable. In the chiller use case, the main work was to create a proxy corresponding to the chiller and a proxy corresponding to the global process. These proxies ensure the right use of the protocols and carry out the mediation operations for the translation of the collected data into a format understandable by the other components [4]. Also, the use of the database allows to keep data longer and thus to manage a form of history. This feature is a real advantage compared to most existing platforms.

**Model execution support**. The execution of the model is performed within a specific docker called by the ML manager. This encapsulation provides a lot of flexibility and decoupling. The model can be easily updated from the cloud. Also, the ML manager collects the necessary data in a dynamic way. Thus, the model always has the data necessary for its execution.

**Model retraining support**. The ML manager continuously collects data that is needed by the model, not only when a prediction is required. It also adds labels as soon as they are available (or, more precisely, computable). This allows to implement continuous learning solutions at the local level or to allow deeper re-training at the cloud level.

**Model versioning**. An important feature of machine learning models is their high dynamicity [12]. They are invalidated by very small changes in the data and must therefore be updated very regularly. The infrastructure we propose at the cloud level allows both to automatically set up the re-training and to keep track of the different model versions.

## VI. CONCLUSION

We presented an architecture and a platform to integrate ML components into a pervasive application and to manage the cycle of these components. The proposed platform is based on the notion of microservices. This allows to host a variety of components that are developed by different teams with different programming skills. In order to smooth up this process, it was important to reduce coupling between components in order to allow parallel developments and independent deployments. Dynamism is brought by service-orientation that allows application development through late composition of independent software components.

The platform is today used in industrial pervasive applications in Schneider Electric, integrated in industrial HMI. The production version is nevertheless reinforced with security mechanisms that are essential requirements today.

REFERENCES

[1] J. Bardin, P. Lalanda, and C. Escoffier. Towards an automatic integration of heterogeneous services and devices, 2010. in Proceedings of the IEEE International Conference on Asian-Pacific Service Computing (APSCC).
[2] C. Becker, S. VanSyckel, and G. Schiele. Ubiquitous information technologies and applications, 2013. Lecture Notes in Electrical Engineering, 214(1).
[3] C. Escoffier, R. S. Hall, and P. Lalanda. ipojo: An extensible service oriented component framework, 2007. in Proceedings of the IEEE International Conference on Service Computing (SCC).
[4] P. Lalanda, L. Bellissard, and R. Balter. Asynchronous mediation for integrating business and operational processes. *IEEE Internet Computing*, 10(1):56–64, 2006.
[5] P. Lalanda, J. McCann, and A. Diaconescu. Principles, design and implementation. undergraduate topics in computer science, 2013. Springer.
[6] P. Lalanda, G. Vega, H. Cervantes, and D. Morand. Architecture and pervasive platform for machine learning services in industry 4.0, 2021. in Proceedings of the International Conference on Pervasive Computing and Communications (PerCom) and other Affiliated Events (PerCom Workshops).
[7] Zhaohui Liu, Hongwei Tan, Duo Luo, Guobao Yu, Jin Li, and Zhenyu Li. Optimal chiller sequencing control in an office building considering the variation of chiller maximum cooling capacity, 2017. In Energy and Buildings, 140:430–442.
[8] F.M. Roth, C. Becker, G. Vega, and P. Lalanda. Xware—a customizable interoperability framework for pervasive computing systems. *Pervasive and mobile computing*, 47, 2018.
[9] W. Shi and S. Dustdar. The promise of edge computing. *Comp.*, 49(5), 2016.
[10] Z. Zheng, O. Chen, C. Fan, N. Guan, A. Vishwanath, D. Wang, and F. Liu. An edge based data-driven chiller sequencing framework for hvac electricity consumption reduction in commercial buildings, 2019. IEEE Transactions on Sustainable Computing, October 2019.
[11] Z. Zheng, Q. Chen, C. Fan, N. Guan, A. Vishwanath, D. Wang, and F. Liu. Data driven chiller sequencing for reducing hvac electricity consumption in commercial buildings, 2018. IACM e-Energy'18, Karlsruhe, Germany.
[12] J. Estublier, G. Vega, P. Lalanda and T. Leveque, "Domain Specific Engineering Environments," 2008 15th Asia-Pacific Software Engineering Conference, Beijing, China, 2008, pp. 553-560,